\documentclass[traditabstract]{aa} 
\usepackage{graphicx}
\usepackage{natbib}

\newcommand{\Msun}{\ensuremath{\,{M}_\odot}}                      
\newcommand{\Rsun}{\ensuremath{\,{R}_\odot}}                      
\newcommand{\Teff}{\ensuremath{T_{\rm eff}}}                      
\newcommand{\Mjup}{\ensuremath{\,{M}_{\rm Jup}}}                  
\newcommand{\Rjup}{\ensuremath{\,{R}_{\rm Jup}}}                  
\newcommand{\Teq}{\ensuremath{T_{\rm eq}^{\,\prime}}}             
\newcommand{\safronov}{\ensuremath{\Theta}}                       
\newcommand{\kms}{\,km\,s$^{-1}$}                                 
\newcommand{\ms}{\,m\,s$^{-1}$}                                   
\newcommand{\mss}{\,m\,s$^{-2}$}                                  
\newcommand{\as}{\ensuremath{^{\prime\prime}}}                    
\newcommand{\am}{\ensuremath{^\prime}}                            
\newcommand{\pjup}{\ensuremath{\,\rho_{\rm Jup}}}                 
\newcommand{\psun}{\ensuremath{\,\rho_\odot}}                     
\newcommand{\mc}[1]{\multicolumn{2}{c}{#1}}
\newcommand{\mcc}[1]{\multicolumn{3}{c}{#1}}

\newcommand{\erc}[3]{\mc{\ensuremath{#1^{+#2}_{-#3}}}}

\usepackage{amsmath,amssymb,graphicx}
\begin{document}
\title{Physical properties and transmission spectrum of the
WASP-80 planetary system from multi-colour photometry}

\titlerunning{Physical properties of WASP-80\,b}

   \author{
          L. Mancini\inst{1} 
          \and
          J. Southworth\inst{2}
          \and
          S. Ciceri\inst{1}
          \and
          M. Dominik\inst{3}
          \and
          Th. Henning\inst{1}
          \and
          U.~G. J{\o}rgensen\inst{4,5}
          \and
          A.~F. Lanza\inst{6}
          \and
          M. Rabus\inst{7,1}
          \and
          C. Snodgrass\inst{8}
          \and
          C. Vilela\inst{2}
          \and
          K.~A. Alsubai\inst{9}
          \and
          V. Bozza\inst{10,11}
          \and
          D.~M. Bramich\inst{12}
          \and
          S. Calchi Novati\inst{13,10}
          \and
          G. D'Ago\inst{10,11}
          \and
          R. Figuera Jaimes\inst{14,3}
          \and
          P. Galianni\inst{3}
          \and
          S.-H. Gu\inst{15,16}
          \and
          K. Harps{\o}e\inst{4,5}
          \and
          T. Hinse\inst{17}
          \and
          M. Hundertmark\inst{3}
          \and
          D. Juncher\inst{4,5}
          \and
          N. Kains\inst{14}
          \and
          H. Korhonen\inst{18,4,5}
          \and
          A. Popovas\inst{4,5}
          \and
          S. Rahvar\inst{19,20}
          \and
          J. Skottfelt\inst{4,5}
          \and
          R. Street\inst{21}
          \and
          J. Surdej\inst{22}
          \and
          Y. Tsapras\inst{21,23}
          \and
          X.-B. Wang\inst{15,16}
          \and
          O. Wertz\inst{22}
          }
       \institute{
   Max Planck Institute for Astronomy, K\"{o}nigstuhl 17, 69117 -- Heidelberg, Germany \\
             \email{mancini@mpia.de}
         \and
   Astrophysics Group, Keele University, Staffordshire, ST5 5BG, UK
         \and
   SUPA, University of St Andrews, School of Physics \& Astronomy, North Haugh, St Andrews, KY16 9SS, UK
         \and
   Niels Bohr Institute, University of Copenhagen, Juliane Maries vej 30, 2100 Copenhagen \O, Denmark
         \and
   Centre for Star and Planet Formation, Geological Museum, {\O}ster Voldgade 5-7, 1350 Copenhagen, Denmark
         \and
   INAF -– Osservatorio Astrofisico di Catania, via S.Sofia 78, 95123 Catania, Italy
         \and
   Instituto de Astrof\'isica, Facultad de F\'isica, Pontificia Universidad Cat\'olica de Chile, Av. Vicu\~na Mackenna 4860,
   7820436 Macul, Santiago, Chile
         \and
   Max-Planck-Institute for Solar System Research, Max-Planck Str. 2, 37191 Katlenburg-Lindau, Germany
         \and
   Qatar Foundation, PO Box 5825, Doha, Qatar
         \and
   Dipartimento di Fisica ``E.R. Caianiello'', University of Salerno, Via Giovanni Paolo II, 84084 Fisciano, Italy
         \and
   Istituto Nazionale di Fisica Nucleare, Sezione di Napoli, Napoli, Italy
         \and
   Qatar Environment and Energy Research Institute, Qatar Foundation, Tornado Tower, Floor 19, P.O. Box 5825, Doha, Qatar
         \and
   Istituto Internazionale per gli Alti Studi Scientifici (IIASS), 84019 Vietri Sul Mare (SA), Italy
         \and
   European Southern Observatory, Karl-Schwarzschild-Stra{\ss}e 2, 85748 Garching bei M\"unchen, Germany
        \and
   Yunnan Observatory, Chinese Academy of Sciences, Kunming 650011, China
         \and
   Key Laboratory for the Structure and Evolution of Celestial Objects, Chinese Academy of Sciences, Kunming 650011, China
         \and
   Korea Astronomy and Space Science Institute, Daejeon 305-348, Republic of Korea
         \and
   Finnish Centre for Astronomy with ESO (FINCA), University of Turku, V{\"a}is{\"a}l{\"a}ntie 20, FI-21500 Piikki{\"o}, Finland
         \and
   Department of Physics, Sharif University of Technology, P.\,O.\,Box 11155-9161 Tehran, Iran
         \and
   Perimeter Institute for Theoretical Physics, 31 Caroline St. N., Waterloo, ON, N2L 2Y5,Canada
         \and
   Las Cumbres Observatory Global Telescope Network, 6740B Cortona Drive, Goleta, CA 93117, USA
         \and
   Institut d'Astrophysique et de G\'eophysique, Universit\'e de Li\`ege, 4000 Li\`ege, Belgium
         \and
   School of Physics and Astronomy, Queen Mary University of London, Mile End Road, London, E1 4NS, UK
         }

\abstract{WASP-80 is one of only two systems known to contain a
hot Jupiter which transits its M-dwarf host star. We present eight
light curves of one transit event, obtained simultaneously using
two defocussed telescopes. These data were taken through the
Bessell $I$, Sloan $g^{\prime}r^{\prime}i^{\prime}z^{\prime}$ and
near-infrared $JHK$ passbands. We use our data to search for
opacity-induced changes in the planetary radius, but find that all
values agree with each other. Our data are therefore consistent
with a flat transmission spectrum to within the observational
uncertainties. We also measure an activity index of the host star
of $\log R^{\prime}_{\rm\,HK} = -4.495$, meaning that WASP-80\,A
shows strong chromospheric activity. The non-detection of
starspots implies that, if they exist, they must be small and
symmetrically distributed on the stellar surface. We model all
available optical transit light curves and obtain improved physical
properties and orbital ephemerides for the system.}


\maketitle

\section{Introduction}
\label{sec:1}

Planetary systems in which the host star is a late-type dwarf are
of particular interest because they have favourable ratios of
planetary mass and radius to those of the star. This makes the
detections of small and low-mass planets easier for the transit
and Doppler methods, respectively. If the planet is transiting and
the parent star is bright, then the planetary atmosphere can be
probed by transmission spectroscopy (e.g.\ GJ\,1214\,b:
\citealp{croll2011,crossfield2011,bean2011,berta2012,colon2013};
GJ\,3470\,b: \citealp{crossfield2013}; GJ\,436\,b:
\citealp{pont2009,gibson2011}), transmission photometry (e.g.\
GJ\,1214\,b:
\citealp{demooij2012,murgas2012,demooij2013,narita2013};
GJ\,3470\,b: \citealp{nascimbeni2013}) and observations of
secondary eclipses (e.g.\ GJ\,1214\,b: \citealp{fraine2013};
GJ\,436\,b: \citealp{stevenson2010,knutson2011}).

Recent analyses of HARPS and {\it Kepler} data suggest that
Neptunes and super-Earths with orbital periods shorter than
50\,days are very abundant around M stars
\citep{bonfils2013,dressing2013}. \citet{bonfils2013} also
established that giant planets have a much lower occurrence rate
for orbital periods in the range 10--100\,d, supporting the idea
that the frequency of giant planets decreases toward less massive
parent stars, irrespective of period \citep{johnson2010}.


Accordingly, only two transiting hot Jupiters have so far been
found orbiting M dwarfs\footnote{Two brown dwarfs are also known
to transit M stars: NLTT\,41135 \citep{irwin2010} and LHS\,6343
\citep{johnson2011}.}. These are Kepler-45\,b (KOI-254;
$R_{\mathrm{p}}=0.999$\Rjup, $M_{\mathrm{p}}=0.500$\Mjup,
\citealp{johnson2012,southworth2012}) and WASP-80\,b
($R_{\mathrm{p}}=0.95$\Rjup, $M_{\mathrm{p}}=0.55$\Mjup,
\citealp{triaud2013}). Whilst Kepler-45 is a distant (333\,pc) and
faint ($V=16.9$) star, WASP-80 is much closer (60\,pc) and
brighter ($V=11.9$), enabling a detailed study of its
characteristics with ground-based facilities. Moreover, due to its
low density and large transit depth ($\sim$3\%), WASP-80\,b is a
very suitable target for transmission spectroscopy and photometry.

Here we present photometric observations of a transit of
WASP-80\,b, observed simultaneously with two telescopes and in
eight different passbands. We use these data to refine the
physical parameters of the planetary system and provide the first
probe of the day-night terminator region of this giant planet by
transmission photometry.

\section{Observations and data reduction}
\label{sec:2}

\begin{table*}
\caption{Details of the transit observations presented in this
work. $N_{\rm obs}$ is the number of observations, $T_{\rm exp}$
is the exposure time, $T_{\rm obs}$ is the observational cadence,
and `Moon illum.' is the fractional illumination of the Moon at
the midpoint of the transit. The aperture sizes are the radii of
the software apertures for the star, inner sky and outer sky,
respectively. Scatter is the r.m.s. scatter of the data versus a
fitted model.} %
\label{tab:01} %
\centering     %
\tiny          %
\setlength{\tabcolsep}{5pt}
\begin{tabular}{lcccrrcccccc}
\hline\hline
Telescope & Date of   & Start time & End time & $N_{\rm obs}$ & $T_{\rm exp}$ & $T_{\rm obs}$ & Filter & Airmass & Moon   & Aperture   & Scatter \\
          & first obs. & (UT)       & (UT)     &                    & (s)           & (s)           &        &    & illum. & radii (px) & (mmag) \\
\hline
DFOSC & 2013 06 16 & 05:35 & 09:44 & 200 & 60 &  75 & Bessel $I$         & $1.33 \rightarrow 1.12 \rightarrow 1.37$ & $ 46\%$ & 16, 38,   60 & 0.49 \\
GROND & 2013 06 16 & 05:00 & 10:50 & 162 & 60 & 120 & Sloan $g^{\prime}$ & $1.38 \rightarrow 1.12 \rightarrow 1.79$ & $ 46\%$ & 30, 90, 120 & 0.80 \\
GROND & 2013 06 16 & 05:00 & 10:50 & 162 & 60 & 120 & Sloan $r^{\prime}$ & $1.38 \rightarrow 1.12 \rightarrow 1.79$ & $ 46\%$ & 35, 100,120 & 0.49 \\
GROND & 2013 06 16 & 05:00 & 10:50 & 162 & 60 & 120 & Sloan $i^{\prime}$ & $1.38 \rightarrow 1.12 \rightarrow 1.79$ & $ 46\%$ & 35, 80, 100 & 0.82 \\
GROND & 2013 06 16 & 05:00 & 10:50 & 162 & 60 & 120 & Sloan $z^{\prime}$ & $1.38 \rightarrow 1.12 \rightarrow 1.79$ & $ 46\%$ & 30, 80, 100 & 1.06 \\
GROND & 2013 06 16 & 05:00 & 10:50 & 523 &  4 &  38 &       $J$          & $1.38 \rightarrow 1.12 \rightarrow 1.79$ & $ 46\%$ &  6, 11,  21 & 4.15 \\
GROND & 2013 06 16 & 05:00 & 10:50 & 523 &  4 &  38 &       $H$          & $1.38 \rightarrow 1.12 \rightarrow 1.79$ & $ 46\%$ &  5, 12,  22 & 3.28 \\
GROND & 2013 06 16 & 05:00 & 10:50 & 523 &  4 &  38 &       $K$          & $1.38 \rightarrow 1.12 \rightarrow 1.79$ & $ 46\%$ &  7, 11,  20 & 5.14 \\

\hline
\end{tabular}
\end{table*}

\begin{table}
\caption{Excerpts of the light curves of WASP-80: this table will
be made available at the CDS. A portion is shown here for guidance
regarding its form and content.}
\label{tab:02} %
\centering     %
\tiny          %
\begin{tabular}{lccrc}
\hline\hline
Telescope    & Filter & BJD (TDB) & Diff. mag. & Uncertainty  \\
\hline
DK 1.54-m  & $I$          & 2456459.732729 &  0.00002 & 0.00055 \\
DK 1.54-m  & $I$          & 2456459.733794 &  0.00029 & 0.00055 \\[2pt]
ESO 2.2-m  & $g^{\prime}$ & 2456459.708924 &  0.00122 & 0.00087 \\
ESO 2.2-m  & $g^{\prime}$ & 2456459.710953 &  0.00079 & 0.00087 \\[2pt]
ESO 2.2-m  & $r^{\prime}$ & 2456459.715126 & -0.00021 & 0.00033 \\
ESO 2.2-m  & $r^{\prime}$ & 2456459.719854 &  0.00084 & 0.00033 \\[2pt]
ESO 2.2-m  & $i^{\prime}$ & 2456459.715126 &  0.00009 & 0.00064 \\
ESO 2.2-m  & $i^{\prime}$ & 2456459.716988 & -0.00059 & 0.00064 \\[2pt]
ESO 2.2-m  & $z^{\prime}$ & 2456459.715126 &  0.00041 & 0.00064 \\
ESO 2.2-m  & $z^{\prime}$ & 2456459.719854 & -0.00109 & 0.00064 \\
\hline
\end{tabular}
\end{table}

A complete transit of WASP-80\,b was observed on 2013 June 16 (see
Table\,\ref{tab:01}), using the DFOSC imager mounted on the 1.54-m
Danish Telescope at ESO La Silla during the 2013 observing
campaign by the MiNDSTEp consortium \citep{dominik2010}. The
instrument has a field of view of 13.7\am$\times$13.7\am\ and a
plate scale of 0.39\as\,pixel$^{-1}$. The observations were
performed through a Bessel $I$ filter and using the
\emph{defocussing} method. The telescope was autoguided and the
CCD was windowed to reduce the readout time. The night was
photometric.

\begin{figure*}
\centering
\includegraphics[width=16.cm]{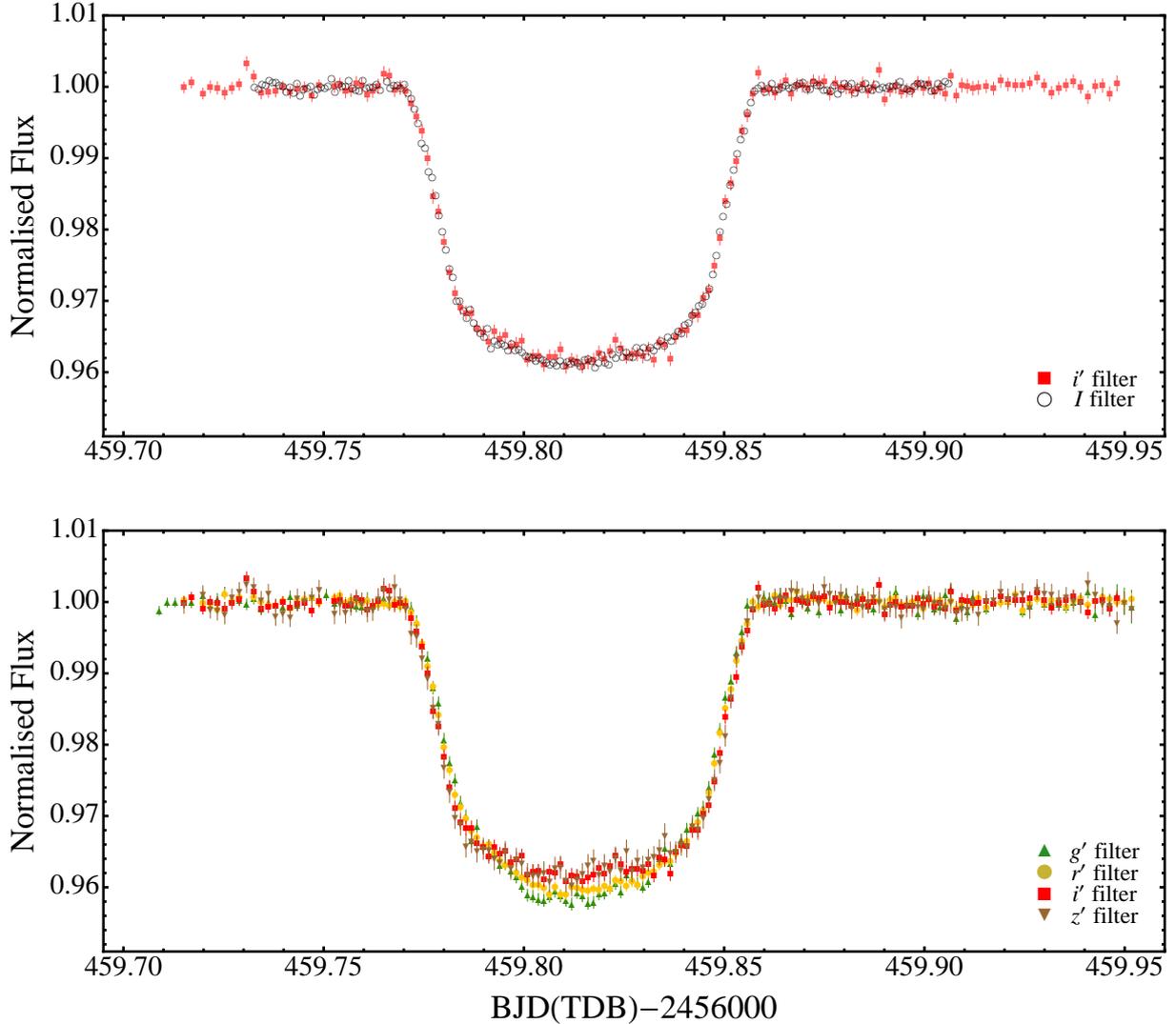}
\caption{Light curves of a transit of WASP-80\,b. \textit{Top
panel}: light curves obtained with the Danish Telescope
(Bessel-$I$ filter) and with GROND (Sloan-$i^{\prime}$),
highlighting the good match between the transit shapes in the two
independent observations. The circles denoting the DK points have
the same size of the corresponding error bars, which have been
suppressed for clarity. \textit{Bottom panel}: light curves
obtained with GROND through four optical filters simultaneously,
showing how the transit shape changes with wavelength.}
\label{Fig:01}
\end{figure*}

The data were reduced using {\sc defot}, an {\sc
idl}\footnote{{\sc idl} is a trademark of the ITT Visual
Information Solutions: {\tt
http://www.ittvis.com/ProductServices/IDL.aspx}} pipeline for
time-series photometry \citep{southworth2009al1}. The images were
debiased and flat-fielded using standard methods, then subjected
to aperture photometry using the {\sc aper}\footnote{{\sc aper} is
part of the {\sc astrolib} subroutine library distributed by NASA
on {\tt http://idlastro.gsfc.nasa.gov}.} task and an optimal
ensemble of comparison stars. Pointing variations were followed by
cross-correlating each image against a reference image. The shape
of the light curve is very insensitive to the aperture sizes, so
we chose those which yielded the lowest scatter. The final light
curve was detrended to remove slow instrumental and astrophysical
trends by fitting a straight line to the out-of-transit data. This
process was simultaneous with the optimisation of the weights of
an ensemble of comparison stars. The final differential-flux light
curve is plotted in Fig.\,\ref{Fig:01}.

The same transit was also observed using the \textbf{G}amma
\textbf{R}ay Burst \textbf{O}ptical and \textbf{N}ear-Infrared
\textbf{D}etector (GROND) instrument mounted on the
MPG\footnote{Max Planck Gesellschaft.}/ESO 2.2-m telescope, also
located at ESO La Silla. GROND is an imaging system capable of
simultaneous photometric observations in four optical (similar to
Sloan $g^{\prime}$, $r^{\prime}$, $i^{\prime}$, $z^{\prime}$) and
three NIR ($J,\, H,\, K$) passbands \citep{greiner2008}. Each of
the four optical channels is equipped with a back-illuminated
$2048 \times 2048$ E2V CCD, with a field of view of $5.4^{\prime}
\times 5.4^{\prime}$ at $0.158^{\prime\prime}\rm{pixel}^{-1}$. The
three NIR channels use $1024 \times 1024$ Rockwell HAWAII-1 arrays
with a field of view of $10^{\prime}\times 10^{\prime}$ at
$0.6^{\prime\prime}\rm{pixel}^{-1}$. The observations were also
performed with the telescope defocussed.


The optical data were reduced as for the Danish Telescope, except
that a quadratic function was used to detrend the light curves.
The NIR data were reduced following the procedure described in
\citet{mancini2013c}. The optical light curves are plotted
superimposed in the bottom panel of Fig.\,\ref{Fig:01} in order to
highlight the differences among the transit depths and light curve
shapes of the simultaneous multi-band observations. The GROND
$i^{\prime}$ and Danish Telescope $I$ light curves are in
excellent agreement. The differential-magnitude light curves are
given in Table\,\ref{tab:02}.

\subsection{Stellar activity measurement}
We obtained a spectrum of the Ca\,II H and K lines on the night of
2012 October 1, using the William Herschel Telescope with the ISIS
grating spectrograph, in order to measure the $\log
R^\prime_{\rm\,HK}$ stellar activity index. With the H2400B
grating we obtained a spectrum covering 375--415\,nm at a
reciprocal dispersion of 0.011\,nm\,pixel$^{-1}$. An exposure time
of 600\,s yielded a continuum signal to noise ratio of
approximately 20 in the region of the H and K lines. The data were
reduced using optimal extraction as implemented in the {\sc
pamela} and {\sc molly} packages \citep{Marsh89pasp} and
calibrated onto the Mt.\ Wilson system using 20 standard stars
from \citet{Vaughan++78pasp}; further details can be found in
Vilela et al.\ (in prep).

\begin{figure}
\centering
\includegraphics[width=\columnwidth]{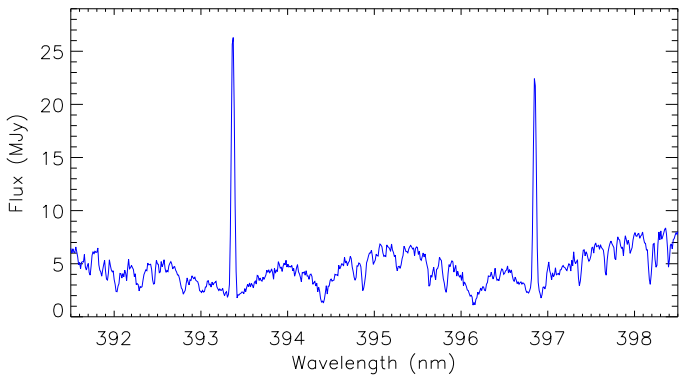}
\caption{Spectrum of WASP-80 in the region of the Ca\,II H and K
lines showing the strong chromospheric emission in the line
cores.} \label{Fig:02}
\end{figure}

The very strong H and K emission lines for WASP-80\,A
(Fig.\,\ref{Fig:02}) yield the emission measure $\log
R^\prime_{\rm\,HK} = -4.495$, which is indicative of high activity
\citep[e.g.][]{Noyes+84apj}. Given this strong chromospheric
emission one might expect to see evidence of spot activity, but
\citet{triaud2013} found no rotational modulation in the SuperWASP
light curves to a limit of $\sim$1\,mmag, and we see no evidence
of spot anomalies \citep[e.g.][]{Tregloan++13mn} in our light
curves. A plausible explanation is that the stellar surface
contains many spots which are too small to noticeably affect
transit light curves, and which are approximately symmetrically
distributed so cause no measurable rotation signal in the
SuperWASP light curves.

This suggestion is in agreement with the conclusions of
\citet{jackson2012} \citep[see also][]{jackson2013}, who analysed
two large samples of low-mass stars ($0.2<M_{\star}/M_{\sun}<0.7$)
in the open cluster NGC\,2516: one sample with measurable
rotational modulation and one without. The two samples coincide on
the colour-magnitude diagram for the cluster, and have the same
rotational velocities and levels of chromospheric activity. This
difference can be explained by the photometrically constant stars
having many small starspots rather than few large ones.

In the case of WASP-80, assuming that the photometric modulation
due to starspots is smaller than 1\,mmag, we can estimate the
maximum deviation of the covering factor from uniformity of a few
$10^{-3}$ of the disc area, by considering that the spot
temperature contrast is probably smaller in cooler stars with
respect to the Sun (see \citealp{berdyugina2005}).

\section{Light-curve analysis}
\label{sec:3}
The light curves were modelled using the {\sc
jktebop}\footnote{The source code of \textsc{jktebop} is available
at: {\tt http:// www.astro.keele.ac.uk/jkt/codes/jktebop.html}}
code \citep[see][and references therein]{southworth2012}, which
represents the star and planet as biaxial spheroids for
calculation of the reflection and ellipsoidal effects and as
spheres for calculation of the eclipse shapes. The main parameters
fitted by {\sc jktebop} are the orbital inclination, $i$, the
transit midpoint, $T_0$, and the sum and ratio of the fractional
radii of the star and planet, $r_{\mathrm{A}}+r_{\mathrm{b}}$ and
$k = r_{\mathrm{b}}/r_{\mathrm{A}}$. The fractional radii are
defined as $r_{\mathrm{A}} = R_{\mathrm{A}}/a$ and $r_{\mathrm{b}}
= R_{\mathrm{b}}/a$, where $a$ is the orbital semimajor axis, and
$R_{\mathrm{A}}$ and $R_{\mathrm{b}}$ are the absolute radii of
the star and the planet, respectively. Each light curve was
modelled separately using the quadratic limb darkening (LD) law.
The linear LD coefficients were fitted to the data whereas the
quadratic LD coefficients were fixed at theoretical values
\citep{claret2004} but perturbed by $\pm0.1$ during the process of
error estimation. We assumed that the orbit was circular
\citep{triaud2013}. We included in the fits the coefficients of a
linear (DFOSC) or quadratic (GROND) polynomial versus time in
order to fully account for the uncertainty in the detrending of
the light curves.

Time-series photometry is unavoidably afflicted by correlated
(red) noise which is not accounted for by standard error
estimation algorithms \citep[e.g.][]{CarterWinn09apj}. The {\sc
aper} algorithm we use to perform aperture photometry also tends
to underestimate the true uncertainties in the relative magnitude
measurements. We therefore rescaled the errorbars as in our
previous works \citep{mancini2013a,mancini2013b,mancini2013c},
first to give a reduced $\chi^2$ of $\chi_{\nu}^{2}=1$ and then
using the $\beta$ approach
\citep[e.g][]{gillon2006,winn2008,gibson2008}.

\subsection{Orbital period determination}
\label{sec:3.1}

\begin{table}
\caption{Times of mid-transit of WASP-80\,b and their residuals
versus a linear orbital ephemeris.}
\label{tab:03} %
\centering     %
\tiny
\begin{tabular}{lrrc}
\hline\hline
Time of minimum    & Cycle & Residual & Reference  \\
BJD(TDB)$-2400000$ & no.   & (d)      &            \\
\hline %
$56054.856812 \pm 0.000135 $ &  -23 &  0.000220 & 1    \\
$56134.620911 \pm 0.000222 $ &    3 & -0.000078 & 1    \\
$56180.638678 \pm 0.000165 $ &   18 & -0.000233 & 1    \\
$56459.814384 \pm 0.000044 $ &  109 &  0.000081 & 2    \\
$56459.814284 \pm 0.000096 $ &  109 & -0.000019 & 3    \\
$56459.814414 \pm 0.000082 $ &  109 &  0.000111 & 4    \\
$56459.814350 \pm 0.000103 $ &  109 &  0.000047 & 5    \\
$56459.814233 \pm 0.000161 $ &  109 & -0.000070 & 6    \\
\hline %
\end{tabular}
\tablefoot{References: (1) \citet{triaud2013}; (2) Danish
telescope (this work); (3) GROND $g^{\prime}$ (this work); (4)
GROND $r^{\prime}$ (this work); (5) GROND $i^{\prime}$ (this
work); (6) GROND $z^{\prime}$ (this work).}
\end{table}

\begin{figure*}
\centering
\includegraphics[width=18.cm]{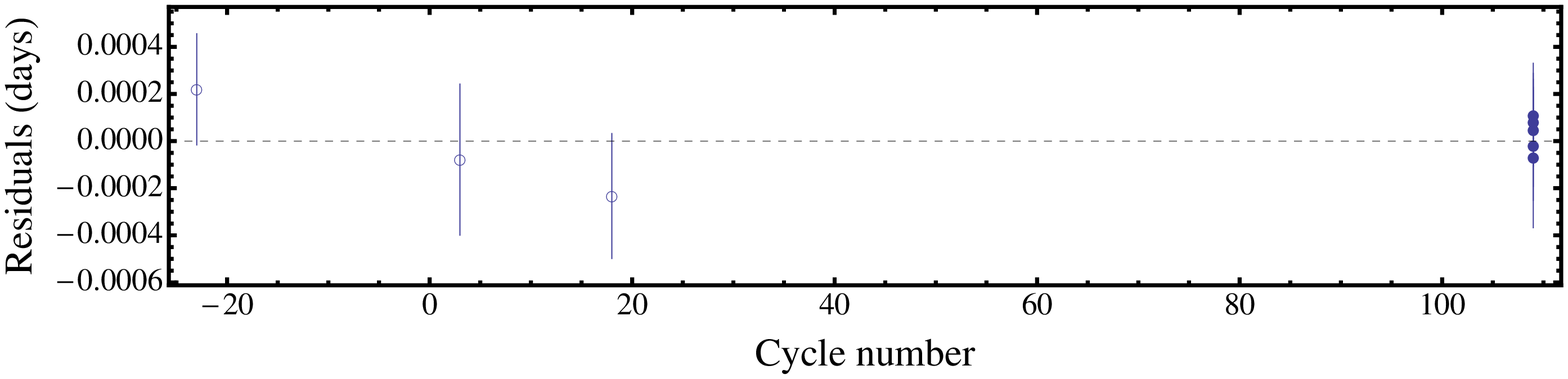}
\caption{Plot of the residuals of the timings of mid-transit of
WASP-80\,b versus a linear ephemeris. Timings based on the
observations obtained by \citet{triaud2013} are plotted using open
circles, while the other timings (this work) are plotted with
filled circles.} \label{Fig:03}
\end{figure*}

We used our photometric data to refine the orbital period of
WASP-80\,b. The transit time for each of our datasets was obtained
by fitting with {\sc jktebop}, and uncertainties were estimated
using Monte Carlo simulations. We also modelled the follow-up
light curves reported in \citet{triaud2013} in order to obtain a
timing for each dataset. All timings were placed on the BJD(TDB)
time system and are summarised in Table\,\ref{tab:03}. The
resulting measurements of transit midpoints were fitted with a
straight line to obtain a final orbital ephemeris:
\begin{equation}
T_{0} = \mathrm{BJD(TDB)} 2\,456\,125.417405(99) +
3.06786144(87)\,E, \nonumber
\end{equation}
where $E$ is the number of orbital cycles after the reference
epoch, which we take to be that estimated by \citet{triaud2013},
and quantities in brackets denote the uncertainty in the final
digit of the preceding number. The quality of fit,
$\chi_{\nu}^2=0.99$, indicates that a linear ephemeris is a good
match to the observations. A plot of the residuals
(Fig.\,\ref{Fig:03}) shows no evidence for systematic deviations
from the predicted transit times. However, the number of observed
transits of this planet is still very low so transit timing
variations cannot be ruled out.

\subsection{Photometric parameters}
\label{sec:3.2}

\begin{figure*}
\centering
\includegraphics[width=18.cm]{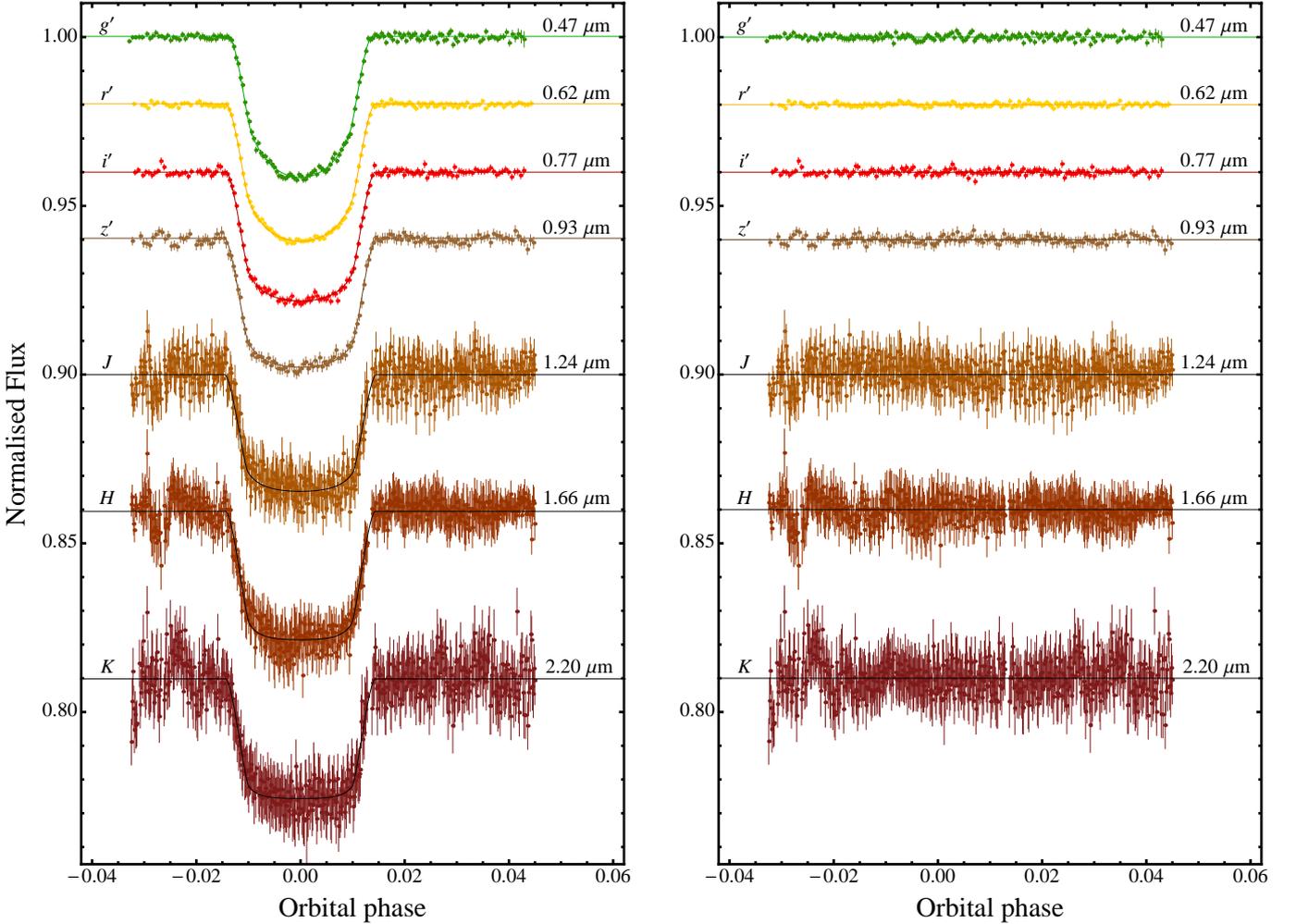}
\caption{\emph{Left-hand panel:} simultaneous optical and NIR
light curves of the transit event of WASP-80\,b observed with
GROND. The {\sc jktebop} best fits are shown as solid lines for
each optical data set. The passbands are labelled on the left of
the figure, and their central wavelengths are given on the right.
\emph{Right-hand panel:} the residuals of each fit.}
\label{Fig:04}
\end{figure*}

\begin{figure*}
\centering
\includegraphics[width=18.cm]{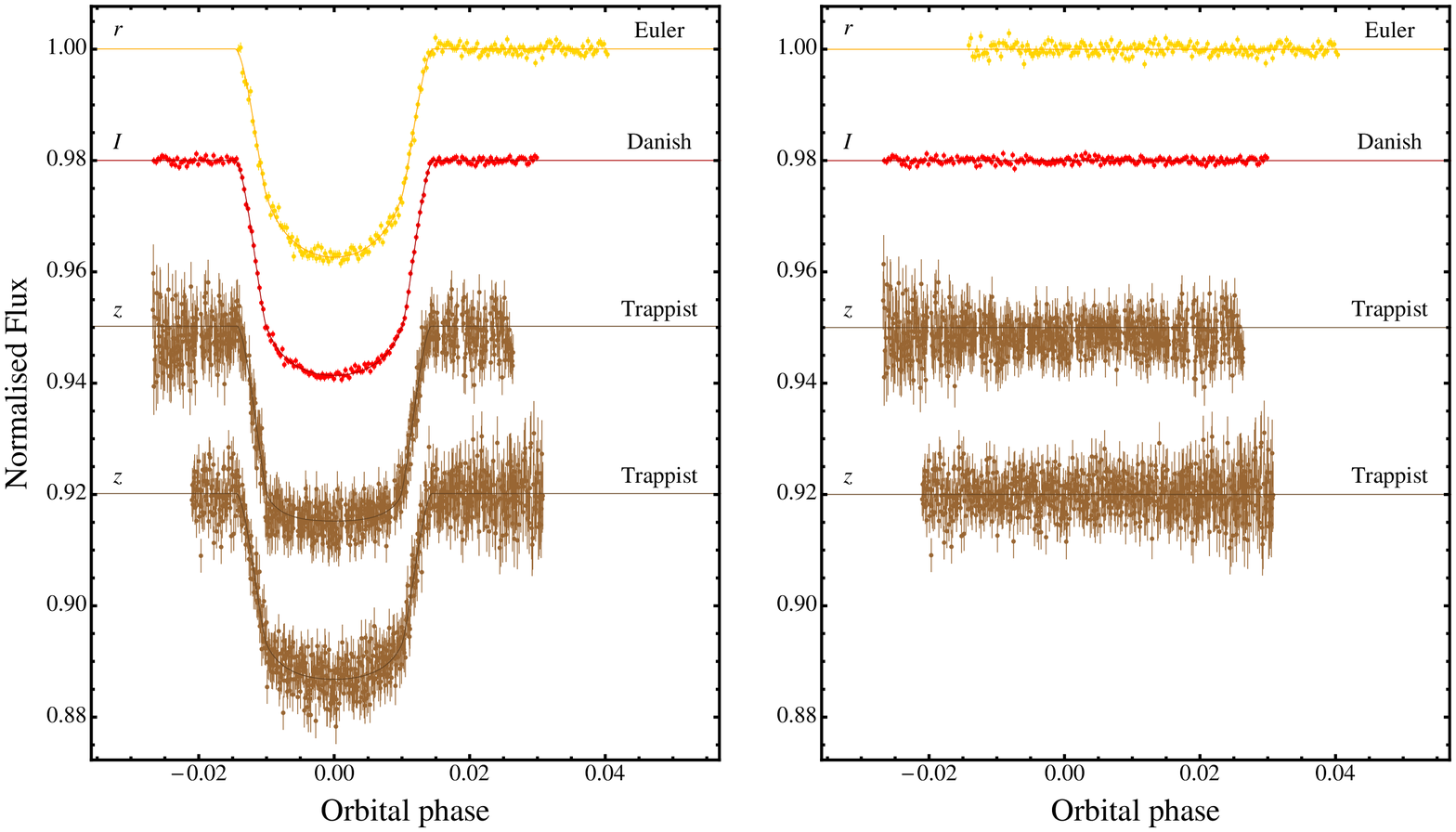}
\caption{\emph{Left-hand panel:} Light curves of transit events of
WASP-80\,b observed in Gunn $r$ with the Euler telescope
\citep{triaud2013}, in Bessell $I$ with the Danish telescope (this
work) and in $z$ with TRAPPIST \citep{triaud2013}. The light
curves are ordered according to central wavelength of the filter
used. The {\sc jktebop} best fits are shown as solid lines for
each optical data set. \emph{Right-hand panel:} the residuals of
each fit.} \label{Fig:05}
\end{figure*}

The GROND light curves and the {\sc jktebop} best-fitting models
are shown in Fig.\,\ref{Fig:04}. A similar plot is reported in
Fig.\,\ref{Fig:05} for the light curves from the Danish Telescope
and from \citet{triaud2013}. The parameters of the fits are given
in Table\,\ref{tab:04}. Uncertainties in the fitted parameters
from each solution were calculated from 3500 Monte Carlo
simulations and by a residual-permutation algorithm
\citep{southworth08}. The larger of the two possible error bars
was retained in each case. The final photometric parameters are
the weighted mean of the results presented in Table\,\ref{tab:04}.
Values obtained by \citet{triaud2013} are also reported for
comparison. Due to their lower quality (see Fig.\,\ref{Fig:04}),
we did not use the GROND-NIR light curves to estimate the final
photometric parameters of WASP-80.

\begin{table*}
\setlength{\tabcolsep}{4.5pt} %
\caption{Parameters of the {\sc jktebop} fits to the light curves
of WASP-80. The final parameters, given in bold, are the weighted
means of the results for the datasets. Results from the discovery
paper are included at the base of the table for comparison.} %
\label{tab:04} %
\centering     %
\tiny          %
\begin{tabular}{llccccc}
\hline\hline
Telescope & Filter & $r_{\mathrm{A}}+r_{\mathrm{b}}$ & $k$ & $i^{\circ}$ & $r_{\mathrm{A}}$ & $r_{\mathrm{b}}$  \\
\hline
Danish\,1.54-m    & Bessel $I$         & $0.09324 \pm 0.00095$ & $0.17135 \pm 0.00099$ & $88.99 \pm 0.23$ & $0.07960 \pm 0.00074$ & $0.01364 \pm 0.00018$ \\[2pt]
MPG/ESO\,2.2-m    & Sloan $g^{\prime}$ & $0.09131 \pm 0.00161$ & $0.17033 \pm 0.00217$ & $89.20 \pm 0.59$ & $0.07802 \pm 0.00128$ & $0.01329 \pm 0.00035$ \\
MPG/ESO\,2.2-m    & Sloan $r^{\prime}$ & $0.09001 \pm 0.00290$ & $0.17041 \pm 0.00175$ & $89.16 \pm 0.50$ & $0.07691 \pm 0.00245$ & $0.01311 \pm 0.00047$ \\
MPG/ESO\,2.2-m    & Sloan $i^{\prime}$ & $0.09239 \pm 0.00167$ & $0.17183 \pm 0.00161$ & $89.10 \pm 0.58$ & $0.07885 \pm 0.00134$ & $0.01355 \pm 0.00033$ \\
MPG/ESO\,2.2-m    & Sloan $z^{\prime}$ & $0.09391 \pm 0.00575$ & $0.17274 \pm 0.00226$ & $89.11 \pm 0.72$ & $0.08007 \pm 0.00091$ & $0.01383 \pm 0.00090$ \\[2pt]
MPG/ESO\,2.2-m    & $J$                & $0.08937 \pm 0.00567$ & $0.16813 \pm 0.00424$ & $90.00 \pm 1.16$ & $0.07651 \pm 0.00476$ & $0.01286 \pm 0.00094$ \\
MPG/ESO\,2.2-m    & $H$                & $0.09393 \pm 0.00365$ & $0.17525 \pm 0.00472$ & $89.34 \pm 0.91$ & $0.07993 \pm 0.00296$ & $0.01401 \pm 0.00076$ \\
MPG/ESO\,2.2-m    & $K$                & $0.09067 \pm 0.00773$ & $0.16383 \pm 0.00809$ & $89.99 \pm 1.29$ & $0.07791 \pm 0.00673$ & $0.01276 \pm 0.00128$ \\[2pt]
Euler\,1.2-m      & Gunn $r$           & $0.09694 \pm 0.00277$ & $0.16726 \pm 0.00270$ & $88.43 \pm 0.45$ & $0.08305 \pm 0.00223$ & $0.01389 \pm 0.00059$ \\[2pt]
Trappist\,0.60-m  & Gunn $z$           & $0.09261 \pm 0.00251$ & $0.17079 \pm 0.00201$ & $88.64 \pm 0.49$ & $0.07910 \pm 0.00204$ & $0.01351 \pm 0.00047$ \\
Trappist\,0.60-m  & Gunn $z$           & $0.09464 \pm 0.00550$ & $0.16285 \pm 0.00230$ & $88.63 \pm 0.87$ & $0.08139 \pm 0.00467$ & $0.01325 \pm 0.00087$ \\
\hline
Final results   & & $\mathbf{0.09283 \pm 0.00058}$ & $\mathbf{0.17058 \pm 0.00057}$ & $\mathbf{88.91 \pm 0.16}$ & $\mathbf{0.07929 \pm 0.00046}$ & $\mathbf{0.01354 \pm 0.00012}$\\
\hline
\citet{triaud2013} & & -- & $0.17126_{-0.00026}^{+0.00031}$ & $89.92_{-0.12}^{+0.07}$ & -- & -- \\
\hline
\end{tabular}
\end{table*}

\section{Physical properties} \label{sec:4}
Following the \emph{Homogeneous Studies} approach \citep[][and
references therein]{southworth2012}, we used the photometric
parameters estimated in the previous section and the spectroscopic
properties of the parent star (velocity amplitude
$K_{\mathrm{A}}=110.9^{+3.0}_{-3.3}$\ms, effective temperature
$T_{\mathrm{eff}}=4145 \pm 100$\,K and metallicity
$\left[\frac{\mathrm{Fe}}{\mathrm{H}}\right]=-0.14 \pm 0.16$) taken from
\citet{triaud2013}, to revise the physical properties of the
WASP-80 system using the {\sc absdim} code.

We iteratively determined the velocity amplitude of the planet
($K_{\mathrm{b}}$) which yielded the best agreement between the
measured $R_{\mathrm{A}}/a$ and \Teff, and those predicted by a
set of theoretical stellar models for the calculated stellar mass
and $\left[\frac{\mathrm{Fe}}{\mathrm{H}}\right]$. The overall
best fit was found over a grid of ages extending from the zero-age
main sequence to a maximum of 5\,Gyr, imposed because WASP-80\,A
shows strong activity indicative of youth (e.g.\
\citealp{west2008}). Statistical errors were propagated by a
perturbation analysis. Systematic errors were estimated by
calculating sets of results using five different sets of
theoretical models
\citep{Claret04aa,Demarque+04apjs,Pietrinferni+04apj,Vandenberg++06apjs,Dotter+08apjs}.
The five models were given equal relative weighting. The resulting
estimates of the physical properties are given in
Table\,\ref{tab:05}. For completeness we also estimated the
physical properties of the WASP-80 system using empirical
calibrations based on detached eclipsing binary systems, instead
of theoretical stellar models, using the method proposed by
\citet{enoch2010} and the calibration equations by
\citet{southworth10}. Table\,\ref{tab:05} shows that the system
properties obtained using the empirical calibration and the
theoretical model sets agree well, except for the models by
\citet{Claret04aa}. This discrepancy can be attributed to the
differences in the \Teff\ scale predicted by the various model
sets \citep[see Fig.\,4 in][]{southworth10}.


Theoretical models prefer a larger age for the system than is
reasonable based on the activity level of the host star. There are
several cases (e.g.\ CoRoT-2 and HD\,189733), where a
planet-hosting star displays a much higher magnetic activity level
than expected for their age \citep{poppenhaeger2013}. A similar
behaviour has also been suggested in the case of Qatar-1
\citep{covino2013}. Several studies pointed out that a close-in
hot-Jupiter can produce different effects on its parent star.
Tidal forces can increase the rotational velocity of the star
\citep{pont09}. The effect of the magnetized stellar wind, which
causes loss of angular momentum, is inhibited by the planet
\citep{lanza2010,cohen2010}. Both processes make the star appear
younger than it is. This situation can be investigated through a
detailed spectroscopic analysis of the host star to refine its
measured atmospheric parameters, and a study of its X-ray
luminosity, which is known to decline with stellar age (e.g.\
\citealp{wright2011}). In the meantime we do not report an
age measurement for the system.

For the final system properties we took the unweighted means of
the four concordant model sets and calculated systematic errorbars
based on the interagreement between them. A comparison between our
final values and those found by \citet{triaud2013} is given in
Table\,\ref{tab:06}.

\begin{table*}
\setlength{\tabcolsep}{3.0pt} %
\flushleft %
\caption{Derived physical properties of the WASP-80 planetary
system using empirical calibrations and each of five sets of
theoretical models.} \label{tab:05}
\begin{tabular}{l r@{\,$\pm$\,}l r@{\,$\pm$\,}l r@{\,$\pm$\,}l r@{\,$\pm$\,}l r@{\,$\pm$\,}l r@{\,$\pm$\,}l}
\hline \hline
\ & \mc{This work} & \mc{This work} & \mc{This work} & \mc{This work} & \mc{This work} & \mc{This work} \\
\ & \mc{(dEB constraint)} & \mc{({\sf Claret} models)} & \mc{({\sf Y$^2$} models)} & \mc{({\sf Teramo} models)} & \mc{({\sf VRSS} models)} & \mc{({\sf DSEP} models)} \\
\hline
$K_{\rm b}$     (\kms) & 122.7   &   2.8    & 126.9   &   1.3    & 122.7   &   2.3    & 122.5   &   2.2    & 124.2   &   1.7    & 123.6   &   1.8    \\
$M_{\rm A}$    (\Msun) & 0.589   & 0.040    & 0.652   & 0.020    & 0.588   & 0.034    & 0.585   & 0.032    & 0.610   & 0.024    & 0.602   & 0.026    \\
$R_{\rm A}$    (\Rsun) & 0.590   & 0.014    & 0.611   & 0.007    & 0.590   & 0.014    & 0.589   & 0.011    & 0.597   & 0.009    & 0.595   & 0.010    \\
$\log g_{\rm A}$ (cgs) & 4.666   & 0.011    & 4.681   & 0.007    & 4.666   & 0.006    & 4.665   & 0.009    & 4.671   & 0.008    & 4.669   & 0.008   \\[2pt]
$M_{\rm b}$    (\Mjup) & 0.557   & 0.030    & 0.596   & 0.020    & 0.557   & 0.030    & 0.555   & 0.025    & 0.571   & 0.022    & 0.566   & 0.023    \\
$R_{\rm b}$    (\Rjup) & 0.981   & 0.024    & 1.015   & 0.014    & 0.981   & 0.023    & 0.979   & 0.020    & 0.993   & 0.016    & 0.989   & 0.017    \\
$\rho_{\rm b}$ (\pjup) & 0.551   & 0.024    & 0.533   & 0.021    & 0.551   & 0.025    & 0.552   & 0.023    & 0.545   & 0.022    & 0.547   & 0.022    \\
\safronov\             & 0.0668  & 0.0024   & 0.0645  & 0.0020   & 0.0668  & 0.0025   & 0.0669  & 0.0023   & 0.0660  & 0.0021   & 0.0663  & 0.0021   \\[2pt]
$a$               (AU) & 0.03464 & 0.00079  & 0.03583 & 0.00037  & 0.03463 & 0.00064  & 0.03457 & 0.00062  & 0.03506 & 0.00047  & 0.03490 & 0.00050  \\
\hline
\end{tabular}
\end{table*}

\begin{table*}
\caption{Final physical properties of the WASP-80 planetary
system, compared with results from \citet{triaud2013}. Separate
statistical and systematic error bars are given for the results
from the current work.} %
\label{tab:06} %
\centering
\begin{tabular}{l l r@{\,$\pm$\,}c@{\,$\pm$\,}l r@{\,$\pm$\,}l r@{\,$\pm$\,}l r@{\,$\pm$\,}l r@{\,$\pm$\,}l}
\hline \hline
\ & \ & \mcc{\bf This work (final)} & \mc{\citet{triaud2013}}  \\
\hline
Stellar mass                        & $M_{\rm A}$    (\Msun) & 0.596     & 0.032     & 0.014   & \erc{0.57}{0.05}{0.05}       \\
Stellar radius                      & $R_{\rm A}$    (\Rsun) & 0.593     & 0.011     & 0.005   & \erc{0.571}{0.016}{0.016}    \\
Stellar surface gravity             & $\log g_{\rm A}$ (cgs) & 4.6678    & 0.0077    & 0.0034  & \erc{4.689}{0.012}{0.013}    \\
Stellar density                     & $\rho_{\rm A}$ (\psun) & \mcc{$2.862 \pm 0.050$}         & \erc{3.117}{0.021}{0.020}    \\[2pt]
Planetary mass                      & $M_{\rm b}$    (\Mjup) & 0.562     & 0.025     & 0.009   & \erc{0.554}{0.030}{0.039}    \\
Planetary radius                    & $R_{\rm b}$    (\Rjup) & 0.986     & 0.020     & 0.008   & \erc{0.952}{0.026}{0.027}    \\
Planetary surface gravity           & $g_{\rm b}$     (\mss) & \mcc{$14.34 \pm  0.46$}         & \erc{15.07}{0.45}{0.42}      \\
Planetary density                   & $\rho_{\rm b}$ (\pjup) & 0.549     & 0.023     & 0.004   & \erc{0.554}{0.030}{0.039}    \\[2pt]
Planetary equilibrium temperature   & \Teq\              (K) & \mcc{$825 \pm  20$}             & \mc{$\sim 800$}              \\
Safronov number                     & \safronov\             & 0.0665    & 0.0023    & 0.0005  & \mc{--}                      \\
Orbital semimajor axis              & $a$               (au) & 0.03479   & 0.00062   & 0.00027 & \erc{0.0346}{0.008}{0.011}   \\
\hline \end{tabular} \end{table*}

\section{Variation of the planetary radius with wavelength}
\label{sec:7}
WASP-80\,b is a good target for studies of the planetary
atmosphere due to the low surface gravity, deep transit, and
bright host star. However, its moderate equilibrium temperature
($T_{\rm eq}^{\,\prime}=825 \pm 20$\,K) indicates that the planet
should belong to the pL class, as suggested by
\citet{fortney2008}. We therefore do not expect a big variation of
the measured planet radius with wavelength. Our GROND data,
however, are very well suited to investigating this possibility as
they cover many passbands.

\begin{figure*}
\centering
\includegraphics[width=18.cm]{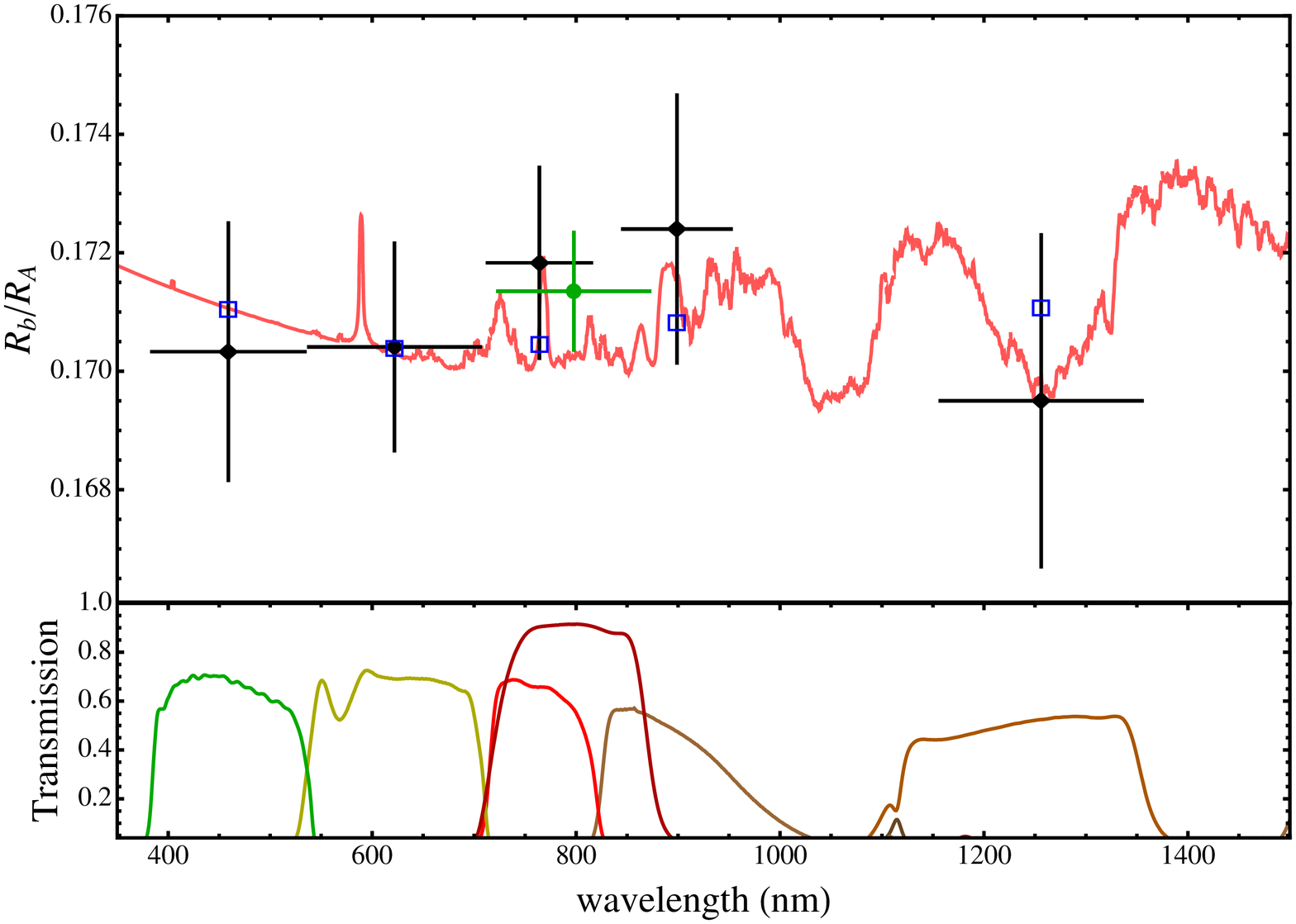}
\caption{Variation of the planetary radius, in terms of
planet/star radius ratio, with wavelength. The black diamonds are
from the transit observations performed with GROND while the green
point is from the same transit observed using the Danish
Telescope. The vertical bars represent the errors in the
measurements and the horizontal bars show the FWHM transmission of
the passbands used. The observational points are compared with a
synthetic spectrum (see text for details). Transmission curves for
the Bessel $I$ filter and the total efficiencies of the GROND
filters are shown in the bottom panel. The blue boxes indicate the
predicted values for the model integrated
over the passbands of the observations.} %
\label{Fig:06}
\end{figure*}

We have measured the ratio of the planetary and stellar radius,
$k$, in the GROND light curves. Fig.\,\ref{Fig:06} shows the
result as a function of wavelength. The vertical errorbars
represent the relative errors in the measurements (i.e.\
neglecting sources of error which affect all light curves
equally), and the horizontal errorbars show the FWHM transmission
of the passbands used. Due to the very large uncertainty, the
values of $k$ measured in the $H$ and $K$ bands were ignored. For
the $J$ band, following \citet{southworth2012al} and
\citet{mancini2013b}, we refitted the data with all parameters
fixed to the final values given in Table 4, with the exception of
k. This approach maximizes the precision of estimations of the
planet/star radius ratio. As our final value for $k$ in the J
band, we got $k = 0.1695 \pm 0.0028$.

The $k$ found for the data from the Danish Telescope is also shown
in green, and is a good match with the results for the GROND
$i^{\prime}$ data. For illustration, Fig.\,\ref{Fig:06} also shows
the predictions from a model atmosphere calculated by
\citet{fortney2010} for a Jupiter-mass planet with gravity
$g_{\mathrm{b}}=10$\,m\,s$^{-2}$, a base radius of $1.25 \,
R_{\mathrm{Jup}}$ at 10\,bar, and $\Teq = 750$\,K. The opacity of
strong-absorber molecules, such as gaseous titanium oxide (TiO)
and vanadium oxide (VO), was removed from the model. Our
experimental points are in agreement with the prominent absorption
features of the model and, being compatible with a flat
transmission spectrum, do not indicate any large variation of the
WASP-80\,b's radius.

\section{Summary and conclusions}
\label{sec:summary}
The WASP-80 system contains a low-mass star which is at the border
of the M spectral class, and a transiting hot Jupiter. It is one
of only two systems containing an M-dwarf and a Jupiter-size
object, the other being the much fainter Kepler-45. The brightness
and small radius of the host star, which means the planetary
transits are deep, and the low surface gravity of the planet make
it a very important object for studying the atmospheric
characteristics of irradiated gas giant planets.

We present eight light curves of one transit event, taken
simultaneously using two telescopes, in the Bessell $I$, Sloan
$g^\prime$, $r^\prime$, $i^\prime$, $z^\prime$, and the near-IR
$JHK$ passbands. We find a good agreement for the ratio of the
planetary to stellar radius determined from our optical light
curves, which means that we do not detect any opacity-induced
changes in planetary radius. We model all available optical
transit light curves and use these results to determine the
physical properties of the system. Our values are consistent with
previous measurements, although our value for the planetary radius
is $1\sigma$ larger than that in the discovery paper.

We also present a spectrum of WASP-80 covering the Ca H and K
lines, which shows strong emission in the line cores. We measure a
chromospheric activity indicator of $\log R^{\prime}_{\rm\,HK} =
-4.495$, which makes WASP-80 one of the most active planet hosts
known. This implies strong magnetic activity and the presence of
starspots, although we see no evidence in our high-precision
photometry for starspot crossing events. If starspots exist on the
surface of WASP-80\,A, they are likely small, numerous and evenly
distributed on the stellar photosphere.

\citet{triaud2013} do not report a conclusive measurement of
$v\,\sin{i_{\star}}$. This is because the one obtained by the
broadening of the spectral lines is incompatible with the one
derived from fitting the Rossiter-McLaughlin effect, suggesting
that the planet's orbital spin could be very inclined. Since the
stellar rotation period is highly uncertain, an estimate of the
stellar age based on gyrochronology is not possible. The age
constraints implied by the stellar models in Sect.\,4 are older
than the strong activity of WASP-80\,A suggests (see
\citealp{pace2013}, Fig.\,1). A possible explanation is that the
stellar activity is enhanced by its planet. However, an estimation
of the stellar age based on theoretical models is very uncertain
because M\,stars evolve very slowly during their main-sequence
lifetime. X-ray observations may help explain this discrepancy.

A detailed characterisation of the atmosphere of WASP-80\,b could
be performed using transmission spectroscopy. We caution that such
investigations should be based on simultaneous observations in
order to avoid complications due to starspot activity.

\begin{acknowledgements}
This paper is based on observations collected with the MPG/ESO
2.2-m and the Danish 1.54-m telescopes, both located at ESO La
Silla, Chile. Operation of the MPG/ESO\,2.2-m telescope is jointly
performed by the Max Planck Gesellschaft and the European Southern
Observatory. Operation of the Danish telescope is based on a grant
to U.G.J.\ by the Danish Natural Science Research Council (FNU).
GROND was built by the high-energy group of MPE in collaboration
with the LSW Tautenburg and ESO, and is operated as a
PI-instrument at the MPG/ESO\,2.2-m telescope. J.S.\ (Keele)
acknowledges financial support from STFC in the form of an
Advanced Fellowship. C.S.\ received funding from the European
Union Seventh Framework Programme (FP7/2007-2013) under grant
agreement no.\ 268421. M.R.\ acknowledges support from FONDECYT
postdoctoral fellowship N$^\circ$3120097. S.-H.G.\ and X.-B.W.\
would like to thank the financial support from National Natural
Science Foundation of China (No.10873031) and Chinese Academy of
Sciences (project KJCX2-YW-T24). O.W.\ thanks the Belgian National
Fund for Scientific Research (FNRS). J.S.\ and O.W.\ acknowledge
support from the Communaut\'{e} francaise de Belgique -- Action de
recherche concert\'{e}es -- Acad\'{e}mie universitaire
Wallonie-Europe. K.A., M.D. and M.H. acknowledge grant
NPRP-09-476-1-78 from the Qatar National Research Fund (a member
of Qatar Foundation).
%
%
The reduced light curves presented in this work will be made
available at the CDS (http://cdsweb.u-strasbg.fr/). The following
internet-based resources were used in research for this paper: the
ESO Digitized Sky Survey; the NASA Astrophysics Data System; the
SIMBAD data base operated at CDS, Strasbourg, France; and the
arXiv scientific paper preprint service operated by Cornell
University.
\end{acknowledgements}

\end{document}